\documentclass[preprint,12pt]{elsarticle}

\usepackage{amssymb}
\usepackage{amsmath}
\usepackage{float,times}
\usepackage{pdflscape}
\usepackage{lscape}
\usepackage{graphicx}
\usepackage{booktabs}
\usepackage{rotating}
\usepackage{booktabs,caption}
\usepackage[flushleft]{threeparttable}
\usepackage{ragged2e}
\usepackage{geometry}
\usepackage{pdfpages}
\usepackage{algorithm}
\usepackage{algpseudocode}

\journal{Energy Economics}

\begin{document}

\begin{frontmatter}

\title{Optimizing Information Asset Investment Strategies in the Exploratory Phase of the Oil and Gas Industry: A Reinforcement Learning Approach}

\author[1,2]{Paulo Barros}
\author[3]{Monica Meireles}
\author[4,5]{Jose Luis Silva}

\affiliation[1]{organization={Petrobras},
            addressline={Av ACM 1113, Itaigara}, 
            city={Salvador},
            postcode={41800-700}, 
            country={Brazil}}
\affiliation[2]{organization={ISCTE Business School},
            addressline={Av Forças Armadas}, 
            city={Lisbon},
            postcode={1649-026}, 
            country={Portugal}}

\affiliation[3]{organization={Universidade de Aveiro (UA)},
            addressline={Campus Universitário de Santiago, 3810-193},
            city={Aveiro},
            postcode={1649-026}, 
            country={Portugal}}
            
\affiliation[4]{organization={Oxaala Technologies},
            addressline={St Dinah Silveira de Queirós, 06}, 
            city={Salvador},
            postcode={40296-160}, 
            country={Brazil}}

\begin{abstract}
Our work investigates the economic efficiency of the prevailing "ladder-step" investment strategy in oil and gas exploration, which advocates for the incremental acquisition of geological information throughout the project lifecycle. By employing a multi-agent Deep Reinforcement Learning (DRL) framework, we model an alternative strategy that prioritizes the early acquisition of high-quality information assets. We simulate the entire upstream value chain—comprising competitive bidding, exploration, and development phases—to evaluate the economic impact of this approach relative to traditional methods. Our results demonstrate that front-loading information investment significantly reduces the costs associated with redundant data acquisition and enhances the precision of reserve valuation. Specifically, we find that the alternative strategy outperforms traditional methods in highly competitive environments by mitigating the "winner's curse" through more accurate bidding. Furthermore, the economic benefits are most pronounced during the development phase, where superior data quality minimizes capital misallocation. These findings suggest that optimal investment timing is structurally dependent on market competition rather than solely on price volatility, offering a new paradigm for capital allocation in extractive industries.
\end{abstract}


\end{frontmatter}

\section{Introduction}
\label{intro}

The global oil and gas sector remains a capital-intensive industry, with exploration investments exceeding \$380 billion between 2015 and 2022 \cite{IEA2023}. A significant portion of this capital, typically 15\% to 20\%, is allocated to information assets such as seismic surveys \cite{Rystad2022}. This expenditure is governed by a fundamental trade-off: acquiring high-fidelity information early in the project lifecycle reduces geological uncertainty but requires substantial upfront capital in a market characterized by high volatility and intense competition for leases \cite{Dixit1994, Cramton2017}.

The industry's standard operating procedure often follows a "ladder-step" strategy, where firms incrementally improve information quality as a project advances through gated phases. While this approach minimizes initial sunk costs, it can lead to sub-optimal outcomes, including redundant data collection, less effective bidding strategies due to information asymmetry, and the misallocation of significant capital during the development phase. Traditional valuation methods, such as Net Present Value (NPV) and standard Real Options Analysis (ROA), often struggle to fully capture the strategic value of information in such dynamic, multi-agent competitive environments \cite{Athey2017, Smith2021}. These static models frequently fail to account for the "information rent" available in competitive bidding or the compound option value of early uncertainty resolution.

Deep Reinforcement Learning (DRL) offers a robust methodological alternative. By modeling the investment process as a stochastic dynamic game, DRL agents can learn optimal policies through trial-and-error interaction with a simulated environment. This approach is uniquely suited to sequential decision-making under uncertainty, where current actions (e.g., purchasing 3D seismic data) influence both future states (e.g., reduced geological variance) and the competitive landscape \cite{Sutton2018}. While DRL has shown promise in optimizing energy trading and operational dispatch \cite{Pereira2022, Baldursson2023}, its application to strategic capital allocation in upstream exploration remains underexplored.

This study makes two primary contributions. First, we develop a novel multi-agent DRL framework that simulates the complete upstream oil and gas value chain, from competitive bidding to production. This model integrates endogenous learning with exogenous market stochasticity. Second, we utilize this framework to evaluate the economic viability of a "front-loaded" investment strategy against the traditional "ladder-step" baseline. We test two core hypotheses: (1) that prioritizing high-quality information acquisition yields superior risk-adjusted returns by mitigating the "winner's curse", and (2) that this advantage is structurally robust across varying market conditions but is amplified by competitive intensity.

Our findings indicate that the RL-optimized strategy significantly outperforms traditional heuristics, particularly in highly competitive auctions where early information precision enhances bidding efficiency. Moreover, we demonstrate that the primary economic value of this strategy is realized during the capital-intensive development phase, where reduced geological uncertainty prevents costly capital misallocation. These results challenge conventional capital budgeting wisdom, suggesting that the optimal timing of information investment is a function of market structure rather than merely price volatility.

The remainder of this paper is organized as follows. Section \ref{litreview} reviews the relevant literature on information economics, investment dynamics, and machine learning in energy. Section \ref{sec:methodology} details the DRL methodology and model specification. Section \ref{sec:data} describes the data and empirical calibration. Section \ref{sec:results} presents the simulation results, followed by a discussion of implications in Section \ref{sec:discussion}. Section \ref{sec:conclusion} concludes.

\section{Theoretical Framework and Related Literature}
\label{litreview}

This research lies at the intersection of information economics, strategic investment under uncertainty, and computational economics. We review the foundational literature in these domains to contextualize our contribution.

\subsection{The Economics of Information and Strategic Valuation}

Information goods are distinct from physical commodities, characterized by non-rivalry and a value proposition tied directly to uncertainty reduction \cite{Shapiro1999, Varian2004}. In the context of natural resource exploration, geological data (e.g., seismic surveys, well logs) serves as a critical information asset. Its economic value is derived from its capacity to resolve uncertainty regarding reserve size and quality, thereby improving the efficiency of subsequent capital deployment \cite{andersen2007value}.

Valuing these assets is non-trivial. While Discounted Cash Flow (DCF) models provide a baseline, they fail to capture the managerial flexibility inherent in exploration projects. Real Options Analysis (ROA) has emerged as a superior framework, treating exploration as a sequence of compound options to acquire, drill, and develop reserves \cite{dias2004valuation, jafarizadeh2015oil}. Within this framework, information is valued as the reduction in the variance of the underlying asset's value distribution \cite{suslick2009uncertainty}. However, standard ROA models are typically single-agent optimizations that ignore the strategic interactions of a competitive market, such as the "winner's curse" in lease auctions \cite{Cramton2017}.

\subsection{Investment Dynamics in the Upstream Sector}

The upstream oil and gas sector is defined by high capital intensity, long investment horizons, and significant exposure to both geological and market risks \cite{ben2019oil}. Investment decisions are influenced by a complex interplay of reserve characteristics, price expectations, and fiscal regimes \cite{berntsen2018determinants, phan2019crude}. A central strategic challenge is the allocation of capital between physical assets (drilling rigs, platforms) and informational assets (seismic data), where the latter serves to de-risk the former \cite{barros2023}.

In competitive bidding environments, information asymmetry plays a critical role. Firms with superior private information can avoid overbidding for low-quality acreage—a phenomenon known as the "winner's curse". The literature suggests that in such settings, the value of private information is non-linear and increases with the number of bidders \cite{Athey2017}. This implies that the optimal investment strategy is not static but depends dynamically on the competitive structure of the market, a nuance often missed by heuristic "ladder-step" strategies.

\subsection{Machine Learning in Economic Decision-Making}

Reinforcement Learning (RL) has emerged as a powerful tool for solving complex, sequential decision-making problems that are analytically intractable. By maximizing a cumulative reward signal through interaction with an environment, RL agents can learn sophisticated policies that approximate optimal behavior in high-dimensional state spaces \cite{Sutton2018}.

In energy economics, RL has been successfully applied to price forecasting \cite{An2019}, power grid optimization, and portfolio management \cite{Feng2018}. Recent work by \cite{Radovic2022} demonstrated the utility of multi-agent DRL in modeling macro-strategic adaptations to the energy transition. However, the application of DRL to the specific micro-economic problem of optimal information acquisition in exploration remains limited. Existing studies have largely focused on operational efficiency or broad corporate strategy \cite{Ghoddusi2019}, leaving a gap in the dynamic optimization of exploration portfolios under competitive pressure.

This study bridges this gap by integrating the theoretical insights of information economics \cite{bakos1999shared} with the computational power of multi-agent DRL. We move beyond static valuation models to provide a dynamic, game-theoretic perspective on investment strategy in the resource sector.

\section{Methodology}
\label{sec:methodology}

We present a multi-agent Deep Reinforcement Learning (DRL) framework designed to model strategic investment decisions in information assets throughout the upstream oil and gas value chain, encompassing competitive bidding, exploration, development, and production \citep{Radovic2022, barros2023}. This approach addresses the inherent limitations of traditional valuation methods, such as Net Present Value (NPV) and Real Options Analysis (ROA), which often treat projects in isolation and struggle to capture the complex, adversarial dynamics of competitive bidding \citep{Dixit1994, Smith2021}. While ROA effectively values managerial flexibility under uncertainty, it typically assumes a single-agent optimizer, failing to account for the strategic "information rent" and the endogenous evolution of competition found in lease auctions \citep{Athey2017, Lange2023}. By leveraging recent advancements in computational economics \citep{Ghoddusi2019, Powell2022}, we model the exploration process as a stochastic dynamic game. We employ Deep Q-Networks (DQN) to solve this game via function approximation, allowing agents to learn optimal policies through trial-and-error interactions where current actions, such as acquiring seismic data, directly influence future state transitions and long-term reward horizons \citep{Sutton2018, Mnih2015}.

\subsection{Deep Reinforcement Learning Framework}

We formulate the investment problem as a Markov Decision Process (MDP) extended to a multi-agent setting. Unlike traditional dynamic programming, which suffers from the curse of dimensionality when modeling multiple competitors and continuous market variables, DRL agents approximate the optimal policy $\pi^*$ through interaction with a stochastic environment \cite{Sutton2018, Bertsekas2019}. 

\begin{figure}[ht]
    \centering    \includegraphics[width=1\textwidth]{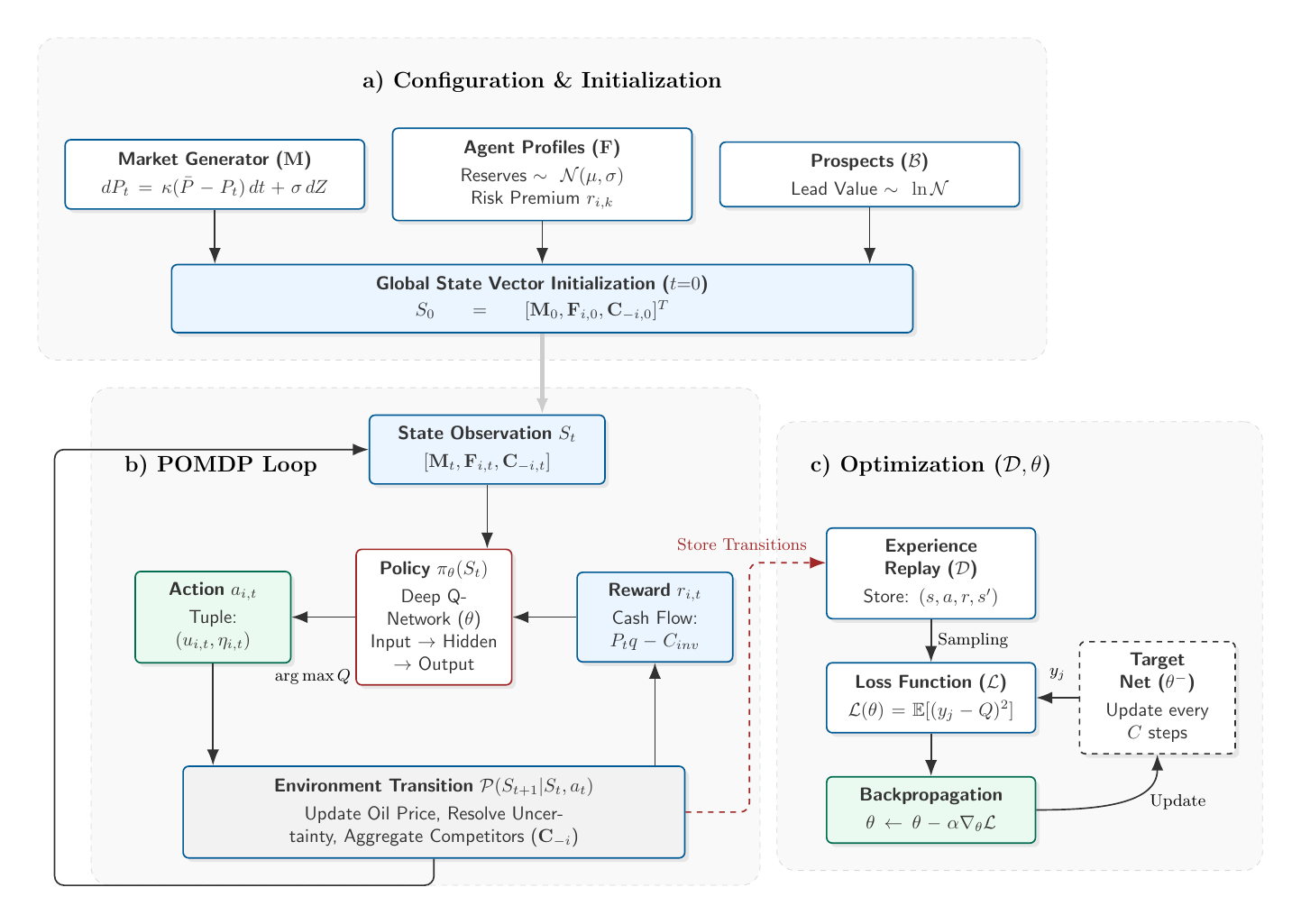} 
    \caption{Deep Reinforcement Learning Architecture for Upstream Investment. Panel (a) illustrates the initialization of market stochasticity ($dP_t$) and agent heterogeneity ($r_{i,k}$). Panel (b) details the POMDP execution loop where the agent observes state $S_t$ (Eq. \ref{eq:state_space}), selects an action $a_t$ (Eq. \ref{fig:rl_actionspace_inv}), and receives a reward $r_t$ (Eq. \ref{fig:rl_reward}). Panel (c) depicts the optimization process using Experience Replay ($\mathcal{D}$) and Target Networks to minimize the temporal difference error.}
    \label{fig:rl_architecture}
\end{figure}

Figure \ref{fig:rl_architecture} delineates the three-stage computational workflow developed to solve this investment game. As shown in Panel (a), the system initializes by generating stochastic market trajectories via the Ornstein-Uhlenbeck process and assigning heterogeneous risk profiles to agents, ensuring diverse competitive scenarios \citep{barros2023}. This configuration feeds into the POMDP execution loop depicted in Panel (b), where the agent perceives the high-dimensional state vector $S_t$ comprising market, firm, and competitor signals (Eq. \ref{eq:state_space}), and outputs an action tuple $a_{i,t}$ (Eq. \ref{fig:rl_actionspace_inv}). The environment responds to this action by transitioning to $S_{t+1}$ and returning a net cash flow reward $r_{i,t}$ (Eq. \ref{fig:rl_reward}), closing the feedback loop required for sequential decision-making.

Panel (c) illustrates the optimization mechanism that runs in parallel to the execution loop. To stabilize the learning of the optimal policy $\pi^*$, we utilize an Experience Replay buffer $\mathcal{D}$ to store transitions $(s, a, r, s')$ and break temporal correlations in the training data \citep{Mnih2015}. The Deep Q-Network parameters $\theta$ are updated by minimizing the temporal difference error against a quasi-static Target Network $\theta^-$, which is refreshed periodically to prevent oscillation during convergence \citep{Li2017}. This architecture allows the agent to learn robust strategies that maximize long-term returns despite the immediate volatility of the exploration phase.

In this framework, an agent $i$ observes a state vector $S_t \in \mathcal{S}$ at discrete time steps $t$, executes an action $a_{i,t} \in \mathcal{A}$, and receives a scalar reward $r_{i,t}$. The environment transitions to a new state $S_{t+1}$ according to the transition probability $\mathcal{P}(S_{t+1} | S_t, a_{i,t}, \mathbf{a}_{-i,t})$, where $\mathbf{a}_{-i,t}$ represents the aggregate actions of competitors. The objective of agent $i$ is to maximize the expected discounted cumulative return $G_t$:

\begin{equation}
G_t = \sum_{k=0}^{\infty} \gamma^k r_{i, t+k}
\end{equation}

where $\gamma \in [0,1]$ is the discount factor reflecting the firm's time preference and capital cost \citep{Ljungqvist2018}. While this general RL objective assumes a constant discount rate, our specific reward function (detailed later in Eq. 5) extends this by incorporating firm-specific risk premiums $r_{i,k}$ to reflect the heterogeneous capital costs of different operators. Specifically, we distinguish between International Oil Companies (IOCs) and National Oil Companies (NOCs). As demonstrated in our empirical calibration (see Section 4.1), these groups exhibit divergent investment behaviors: IOCs typically operate under stricter capital constraints driven by shareholder returns, resulting in higher risk premiums, whereas NOCs often prioritize long-term reserve replacement and resource sovereignty, frequently maintaining higher reserves-to-production ratios \citep{barros2023}.






To effectively handle the high-dimensional and continuous state space inherent in energy markets comprising stochastic oil prices, volatility, and competitor actions, we employ a Deep Q-Network (DQN) \citep{Mnih2015}. Unlike traditional Q-learning, which relies on a tabular representation infeasible for continuous domains, DQN approximates the optimal action-value function $Q^*(s, a)$ using a non-linear function approximator, specifically a deep neural network parameterized by weights $\theta$. The network seeks to satisfy the Bellman optimality equation:

\begin{equation}
Q(s, a; \theta) \approx Q^*(s, a) = \mathbb{E} \left[ r_{i,t} + \gamma \max_{a'} Q(S_{t+1}, a'; \theta) \mid S_t=s, a_{i,t}=a \right]
\label{eq:dqn_bellman}
\end{equation}

To ensure training stability and convergence, we minimize the temporal difference (TD) error using two critical mechanisms: \textit{Experience Replay} and a \textit{Target Network}. The network weights $\theta$ are optimized by minimizing the loss function $L(\theta)$ at each training iteration $j$:

\begin{equation}
L_j(\theta_j) = \mathbb{E}_{(s,a,r,s') \sim \mathcal{D}} \left[ \left( r + \gamma \max_{a'} Q(s', a'; \theta_j^-) - Q(s, a; \theta_j) \right)^2 \right]
\label{eq:dqn_loss}
\end{equation}

where $\mathcal{D}$ is the experience replay buffer that stores past transitions $e_t = (s_t, a_t, r_t, s_{t+1})$ to break temporal correlations in the training data \citep{Li2017}. The term $\theta^-$ denotes the parameters of the quasi-static target network, which are held constant for a fixed number of steps and periodically updated ($\theta^- \leftarrow \theta$) to mitigate the moving target problem and prevent oscillation \citep{Sutton2018}.

\subsection{Model Specification}

We formalize the upstream value chain as a sequential investment game with incomplete information, modeled as a Partially Observable Markov Decision Process (POMDP). This formulation is essential because agents do not have direct access to the ground truth of the geological formations. Therefore, the agents must make capital allocation decisions based on probabilistic beliefs derived from acquired information assets \citep{Littman1994, Yang2018}. The game proceeds through four distinct, gated phases that enforce a strict temporal structure on capital deployment, beginning with the Bidding phase ($\phi=1$) where firms compete for exploration rights in a blind auction. In this initial stage, the incomplete information is twofold: agents are uncertain about the true value of the reservoir and are unaware of competitors private valuations, creating a strategic environment prone to the "winner's curse" \citep{Cramton2017}.

Upon securing rights, the process advances to the Exploration phase ($\phi=2$), which focuses on uncertainty resolution. Therefore, agents invest in seismic acquisition ($\eta$) to receive noisy observations of the hidden geological state, and the transition to the next phase is contingent on these signals reducing the variance of the reserve estimate below a workable threshold. This is followed by the Development phase ($\phi=3$), characterized by irreversible, capital-intensive infrastructure build-out. Decisions at this juncture rely heavily on the belief state constructed during exploration, as valuation errors in the previous phase lead to significant capital misallocation. Finally, the lifecycle concludes with the Production phase ($\phi=4$), where revenue is realized based on stochastic market prices. This sequential structure mirrors the "ladder-step" investment logic prevalent in the industry, where capital is deployed incrementally as geological risk diminishes, but introduces the flexibility for RL agents to deviate and "front-load" information when the expected value of uncertainty reduction exceeds the option value of waiting \citep{barros2023, Dixit1994}.

\paragraph{State Space Representation.} Let $N$ be the set of competing firms. The global state vector at time $t$, $S_t$, is defined as a concatenation of market dynamics, firm-specific states, and competitor aggregates:

\begin{equation}
S_t = [\mathbf{M}_t, \mathbf{F}_{i,t}, \mathbf{C}_{-i,t}]^T
\label{eq:state_space}
\end{equation}

The state vector comprises three distinct components, beginning with the Market State (${M}_t$), which encapsulates the stochastic macroeconomic environment including the oil price $P_t$, following a mean-reverting Ornstein-Uhlenbeck process, price volatility $\sigma_t$, and global demand $D_t$, effectively capturing the external shocks that drive industry-wide investment cycles \citep{Pindyck1999}. The second component, the Firm State (${F}_{i,t}$), tracks the agent's internal status, specifically the current operational phase $\phi_{i,t}$, the volume of proven reserves $R_{i,t}$, and, crucially, the precision of acquired information $\mathcal{I}_{i,t}$. This variable $\mathcal{I}_{i,t}$ distinguishes between low-fidelity (e.g., 2D seismic) and high-fidelity (e.g., 3D/4D seismic) data, serving as a primary lever for strategic differentiation \citep{moody1999measuring}. Finally, the Competitor State (${C}_{-i,t}$) models strategic interdependence without exploding the state space dimensionality by aggregating the observables of the remaining $N-1$ firms. This component tracks total industry investment and total production, allowing the agent to infer competitive intensity and potential "winner's curse" scenarios in real-time \citep{Yang2018, Radovic2022}.

\paragraph{Action Space and Investment Strategies.} The action space $\mathcal{A}$ is rigorously defined to capture the fundamental managerial trade-off between the velocity of capital deployment and the precision of geological intelligence. At each discrete decision node, the agent selects a composite action tuple $a_{i,t} = (u_{i,t}, \eta_{i,t})$.

\begin{equation}
a_{i,t} = (u_{i,t}, \eta_{i,t})
\label{fig:rl_actionspace_inv}
\end{equation}

Here, $u_{i,t} \in \{0, 1\}$ represents the binary managerial decision to essentially exercise a real option: either to proceed to the next operational phase (submitting a binding bid or commissioning a drilling rig) or to defer investment to preserve option value. Simultaneously, the agent must select the quality of the information asset $\eta_{i,t} \in \{\eta_{low}, \eta_{med}, \eta_{high}\}$, which dictates the discrete level of investment in data acquisition technologies, such as choosing between sparse 2D seismic lines versus high-fidelity 3D wide-azimuth surveys. While selecting a higher $\eta$ incurs exponentially greater upfront costs, it provides a commensurate reduction in geological uncertainty, thereby sharpening the agent's valuation of the asset and lowering the effective risk premium associated with the project. Crucially, the model incorporates the principle of avoidable costs. If the agent chooses to defer or abandon a prospect ($u_{i,t}=0$), the model nullifies the cost of information acquisition and prevents the accrual of sunk costs for inactive projects \citep{Dixit1994}.

We contrast two distinct strategic frameworks to evaluate the efficacy of this decision process. The first is the Standard Ladder Strategy (SLS), a heuristic baseline mirroring conservative industry norms where information investment $\eta_{i,t}$ increases monotonically with phase maturity ($\eta \uparrow$ as $\phi \uparrow$). In this risk-averse approach, high-cost data is only acquired after initial uncertainties are resolved, minimizing capital exposure in early, high-risk phases \citep{bratvold2010making}. The second is the RL-Optimized Strategy (RLOS), where the agent endogenously learns the optimal information timing $(\eta^*_{i,t})$ through trial and error. This flexible approach allows for the strategic "front-loading" of high-quality information ($\eta_{high}$ at $\phi=1$), effectively paying an information premium to mitigate the "winner's curse" in competitive auctions and secure a dominant informational position \citep{Athey2017, Hong2023}.






\paragraph{Reward Structure and Market Dynamics.} The immediate reward function $r_{i,t}$ serves as the primary reinforcement signal and is defined as the net cash flow generated at time $t$ to guide the agent toward value-maximizing behaviors. Unlike static valuation methods the Reinforcement Learning agent seeks to maximize the discounted sum of these flows dynamically by continuously adapting to new information.

\begin{equation}
r_{i,t} = \delta_{\phi, 4} \cdot [P_t \cdot q_{i,t} - C_{opr}] - C_{inv}(u_{i,t}) - C_{info}(\eta_{i,t})
\label{fig:rl_reward}
\end{equation}

The equation components function such that $\delta_{\phi, 4}$ acts as a binary indicator for the production phase while $q_{i,t}$ represents the production volume and $C_{opr}$ denotes the operational cost. The term $C_{inv}$ captures the physical capital expenditure contingent on the binary investment decision $u_{i,t}$ and $C_{info}$ represents the cost of information acquisition which scales with the chosen precision $\eta_{i,t}$. The value of information is therefore endogenous as it is considered positive only if the reduction in geological uncertainty leads to a superior investment decision that yields returns exceeding the upfront cost. To strictly enforce economic realism the oil price $P_t$ is modeled to evolve according to a discretized Ornstein-Uhlenbeck mean-reverting process that has been calibrated to historical market data \citep{Pindyck1999}:

\begin{equation}
dP_t = \kappa (\bar{P} - P_t) dt + \sigma_P P_t dZ_t
\end{equation}

This stochastic process is governed by the mean reversion speed $\kappa$ and the long-run equilibrium price $\bar{P}$ while $dZ_t$ represents a Wiener process driving random shocks. These dynamics ensure the agent is exposed to realistic price cycles and volatility $\sigma_P$ during the 25-year simulation horizon which effectively trains the policy to remain robust during periods of market instability.



\subsection{Empirical Implementation}

The model instantiates $N=10$ heterogeneous agents initialized with stochastic financial profiles derived from major offshore operators, including Shell, Exxon, and Petrobras, selected based on historical investment rankings from 2001 to 2021 \citep{barros2023}. These top ten firms account for approximately 40\% of global offshore investments, ensuring the model captures the structural dichotomy between International Oil Companies (IOCs) and National Oil Companies (NOCs), particularly regarding reserve-to-production ratios and capital constraints \citep{Rystad2022}. The simulation horizon is discretized into annual steps corresponding to industry-standard project lifecycles: Bidding ($T_{bid}=1$), Exploration ($T_{exp}=5$), Development ($T_{dev}=7$), and Production ($T_{prod}=25$), with discount factors adjusted dynamically to reflect the phase-specific risk premiums observed in the empirical data.

The training protocol adopts a rigorous curriculum learning approach comprising three distinct stages to ensure robust policy convergence. Initially, agents engage in Self-Play (SP), where they train against clones using an $\epsilon$-greedy exploration strategy that decays from $1.0$ to $0.1$ over the first 50\% of episodes to establish fundamental value evaluations. Subsequently, the agents undergo League Training (LT), where they compete against a pool of fixed "standard strategy" policies—modeled to learn the exploitation of sub-optimal competitors and refine their bidding logic against static baselines. Finally, in the evaluation phase, the optimal hyperparameters identified via grid search ($\alpha=0.01, \gamma=0.50, \epsilon=0.9$) are locked, and performance is assessed using bootstrapping on 10,000 Monte Carlo episodes to ensure statistical significance of the Risk-Adjusted Return on Capital (RAROC) \citep{Efron1994}.

\section{Data and Empirical Calibration}
\label{sec:data}

This section presents a comprehensive empirical analysis of the dataset used to calibrate the reinforcement learning environment. The data captures both firm-level characteristics and global market conditions that shape investment behavior in the offshore oil and gas industry. Our dataset spans from 2001 to 2021, integrating financial reports (F-20 forms) and market tickers sourced from the EIKON platform to ensure empirical consistency. The database is structured across three primary dimensions: firm-specific operational metrics including daily production and reserves, corporate investment patterns segmented by upstream and exploration expenditures, and global market conditions defined by stochastic oil prices and volatility indices calculated via the Levi exponent \citep{barros2023}.

\subsection{Firm-Level Investment Patterns}

We compile data for the top-ten offshore oil and gas investors based on annual capital expenditures. Table \ref{tab:firm_vars} provides a detailed summary of investment, production, and reserve variables, revealing substantial heterogeneity between companies. Shell leads in both mean and maximum investments, with its peak expenditure of \$79.9 billion representing a significant outlier driven by a major asset acquisition strategy in 2016, followed by Exxon and Petrobras, while Equinor, Rosneft, and ONGC exhibit lower mean investment levels despite holding vast reserve bases.

\begin{table}[ht]
\centering
\caption{Descriptive Statistics for Top-Ten Offshore Investors (2001–2021)}
\label{tab:firm_vars}
\resizebox{\textwidth}{!}{%
\begin{tabular}{lccccccccccc}
\toprule
&\multicolumn{3}{c}{\textbf{Investment (\$B USD)}}&&\multicolumn{3}{c}{\textbf{Production (MBOE/d)}}&&\multicolumn{3}{c}{\textbf{Reserves (BBOE)}}\\ 
\cmidrule{2-4}\cmidrule{6-8}\cmidrule{10-12}
\textbf{Firm} & Min & Mean & Max && Min & Mean & Max && Min & Mean & Max\\ 
\cmidrule{2-4} \cmidrule{6-8} \cmidrule{10-12}
Shell & 11.8 & 27.0 & 79.9 && 2.9 & 3.7 & 5.3 && 9.1 & 12.3 & 14.3 \\
Exxon & 12.3 & 25.1 & 42.5 && 3.7 & 4.1 & 4.5 && 11.3 & 20.2 & 25.3 \\
Petrobras & 6.5 & 22.2 & 48.1 && 1.3 & 2.3 & 2.8 && 8.8 & 11.3 & 13.1 \\
Chevron & 7.4 & 20.9 & 41.9 && 2.5 & 2.7 & 3.1 && 10.6 & 11.4 & 12.1 \\
BP & 11.2 & 20.1 & 36.6 && 3.2 & 3.6 & 4.0 && 16.4 & 18.0 & 19.9 \\
Total & 7.7 & 17.4 & 34.4 && 2.2 & 2.5 & 3.0 && 10.5 & 11.3 & 12.7 \\
Cnooc & 1.5 & 13.1 & 31.1 && 0.3 & 0.9 & 1.6 && 1.8 & 3.5 & 5.7 \\
Equinor & 2.3 & 12.4 & 23.8 && 1.0 & 1.8 & 2.1 && 4.3 & 5.2 & 6.2 \\
Rosneft & 0.4 & 9.0 & 16.9 && 0.3 & 3.2 & 5.8 && 1.4 & 24.4 & 44.9 \\
Ongc & 3.2 & 8.3 & 12.1 && 0.6 & 0.7 & 0.9 && 4.5 & 7.7 & 10.5 \\
\bottomrule
\end{tabular}}
\begin{tablenotes}
\small
\item \textit{Note:} Values are annual aggregates. MBOE/d: Million Barrels of Oil Equivalent per day. BBOE: Billion Barrels of Oil Equivalent. ExxonMobil is abbreviated as Exxon, and Equinor represents Statoil.
\end{tablenotes}
\end{table}

\begin{figure}[ht]
    \centering
    \includegraphics[width=1.0\textwidth]{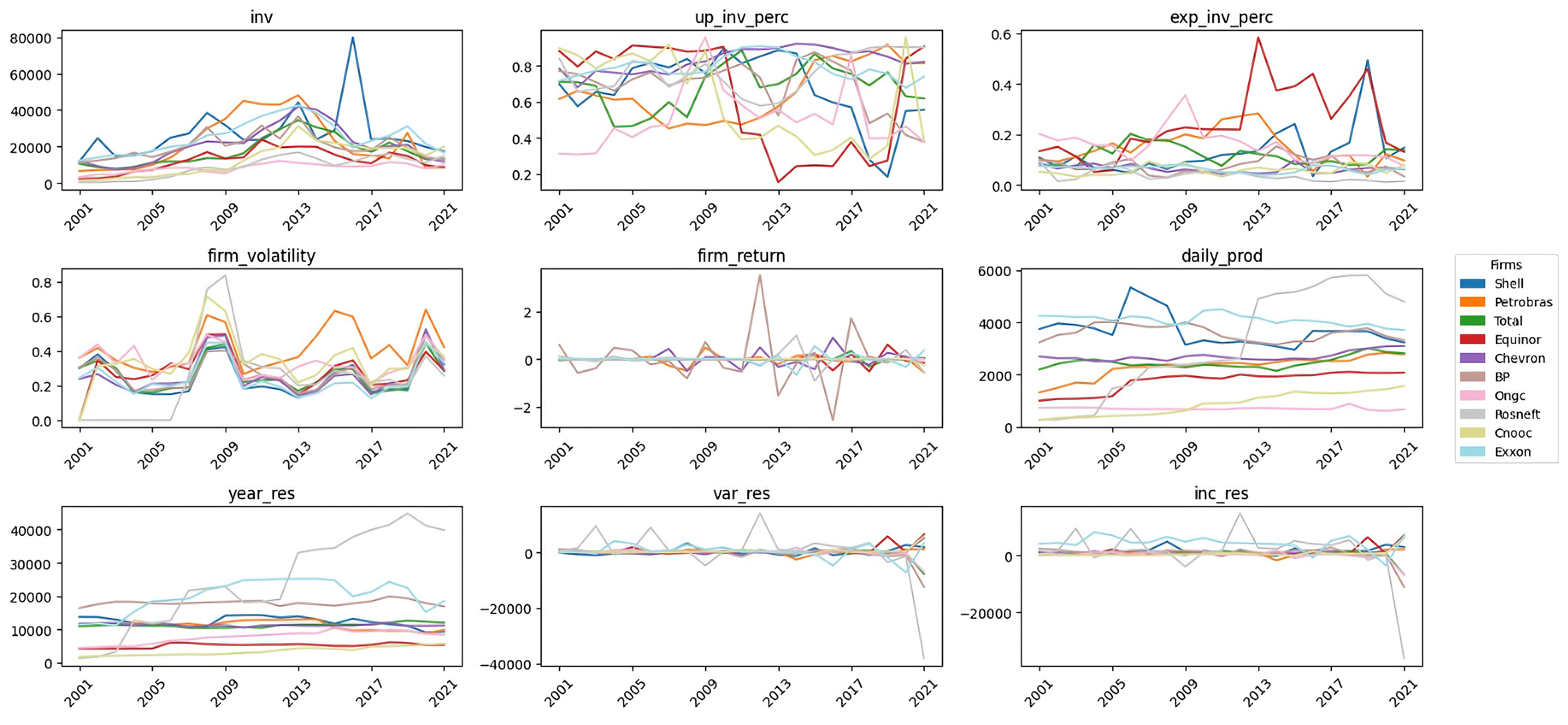}
    \caption{\textbf{Historical Data for Firm and Market Variables.} Historical trends (2001--2021) for key investment, financial, and production variables for the top ten offshore investors: Shell, Petrobras, Total, Equinor, Chevron, BP, ONGC, Rosneft, CNOOC, and Exxon.
    The panels display the evolution of key variables, arranged as follows: 
    (Top-Left) Investment (\textit{inv}),
    (Top-Center) Upstream Investment Percentage (\textit{up\_inv\_perc}),
    (Top-Right) Exploration Investment Percentage (\textit{exp\_inv\_perc}), 
    (Middle-Left) Firm Volatility (\textit{firm\_volatility}),
    (Middle-Center) Firm Return (\textit{firm\_return}),
    (Middle-Right) Daily Production (\textit{daily\_prod}), 
    (Bottom-Left) Yearly Reserves (\textit{year\_res}), 
    (Bottom-Center) Variation in Reserves (\textit{var\_res}), and 
    (Bottom-Right) Reserve Increment (\textit{inc\_res}).
    }
    \label{fig:fig04}
\end{figure}

Figure \ref{fig:fig04} illustrates the historical trends for these firms, showing divergent investment strategies between National Oil Companies (NOCs) and International Oil Companies (IOCs). The investment panels (a, b, c) highlight the scale and allocation of capital expenditure, where Rosneft exhibited the most significant growth, advancing from the lowest-ranked producer in 2001 to the highest by 2013 as shown in the daily production trends (f). This growth was driven by low production costs in the Siberia basin, which allowed for rapid monetization of reserves. Conversely, majors such as Shell, Exxon, and BP experienced stable or declining reserve bases during the same period (g), which required aggressive capital deployment to maintain production levels.

The structural differences between these firm types are quantified by their reserve lifespans. On average, the proven reserves-to-production ratio for NOCs is 5.85 years, compared to 4.4 years for IOCs. This disparity highlights a critical strategic pressure on IOCs: with a lower reserve cushion, these firms face an urgent need for reserve replacement, requiring either increased exploration investments or a managed reduction in production to maintain the balance ratios. The investment distribution reflects this reality, favoring IOCs (approximately 67\%) over NOCs (33\%), even as reserves are more evenly distributed (60\% IOCs vs 40\% NOCs). This empirical evidence validates the initialization of heterogeneous agents in our simulation, where risk premiums are adjusted to reflect these divergent capital constraints.

\subsection{Market Environment and Scenario Generation}

The global O\&G market environment incorporates a comprehensive set of exogenous variables defined in the environment vector $E$, which tracks total industry investments ($tot\_inv$), explicitly segmented into upstream ($tot\_inv\_up$) and exploration ($tot\_inv\_exp$) capital expenditures \citep{barros2023}. Additionally, the environment monitors global operational metrics including production levels ($tot\_prod$), proven reserves ($tot\_res$), and reserve replacement rates ($tot\_inc\_res$), alongside macroeconomic drivers such as oil demand and price volatility. Consistent with the methodology outlined in Section \ref{sec:methodology}, we employ a discretized Ornstein-Uhlenbeck mean-reverting stochastic process for oil prices \citep{Pindyck1999}, calibrated to historical data:
\begin{equation}
dP_t = \kappa(\bar{P} - P_t)dt + \sigma_P P_t dZ_t
\end{equation}
where $\kappa$ represents the speed of mean reversion, $\bar{P}$ is the long-run equilibrium price, and $\sigma_P$ captures price volatility.

We have calibration the model using daily Brent crude spot prices from January 2001 to December 2021. This empirical analysis yields a mean reversion speed $\kappa = 0.38$ and a long-run equilibrium price $\bar{P} = \$65.40$. Notably, the volatility parameter $\sigma_P = 0.28$ was derived using the Levi exponent rather than standard deviation, providing a more robust measure of the heavy-tailed risk distributions characteristic of energy markets \citep{barros2023}. This specification effectively captures the essential dynamics of commodity prices while allowing for persistent volatility shocks \citep{Etheridge2010}.

The scenario configurations, generated by the \textit{ScenarioGenerator} class, introduce dynamic macroeconomic conditions that shape investment outcomes. As illustrated in Figure \ref{fig:fig06}, the generator produces distinct market trajectories—classified as "Resilient," "Neutral," and "Heat" scenarios—which vary by demand growth and price stability. This approach ensures agents are trained against a diverse range of potential futures, preventing overfitting to a single market regime.

\begin{figure}[ht]
    \centering
    \includegraphics[width=0.9\textwidth]{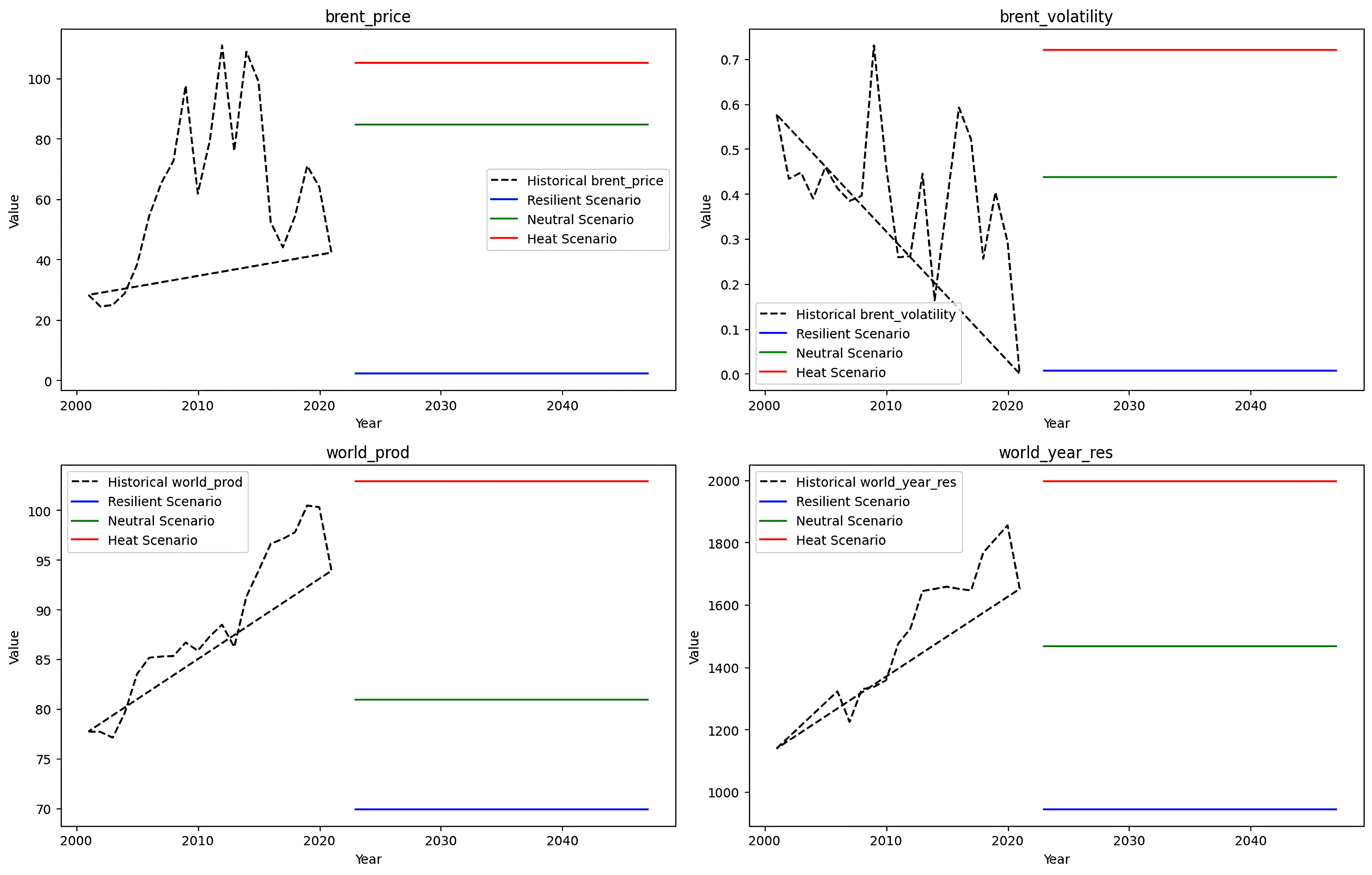}
    \caption{\textbf{Future Scenarios for O\&G Game.} 
    The panels display historical and projected values for key market variables, arranged as follows: 
    (Top-Left) Brent Crude Oil Price (\textit{brent\_price}); 
    (Top-Right) Price Volatility (\textit{brent\_volatility}); 
    (Bottom-Left) Global Production Levels (\textit{world\_prod}); and 
    (Bottom-Right) Global Yearly Reserves (\textit{world\_year\_res}).}
    \label{fig:fig03}
\end{figure}

\subsection{Bid Generation and Firm Profiling}

\begin{figure}[ht]
    \centering
    \includegraphics[width=0.9\textwidth]{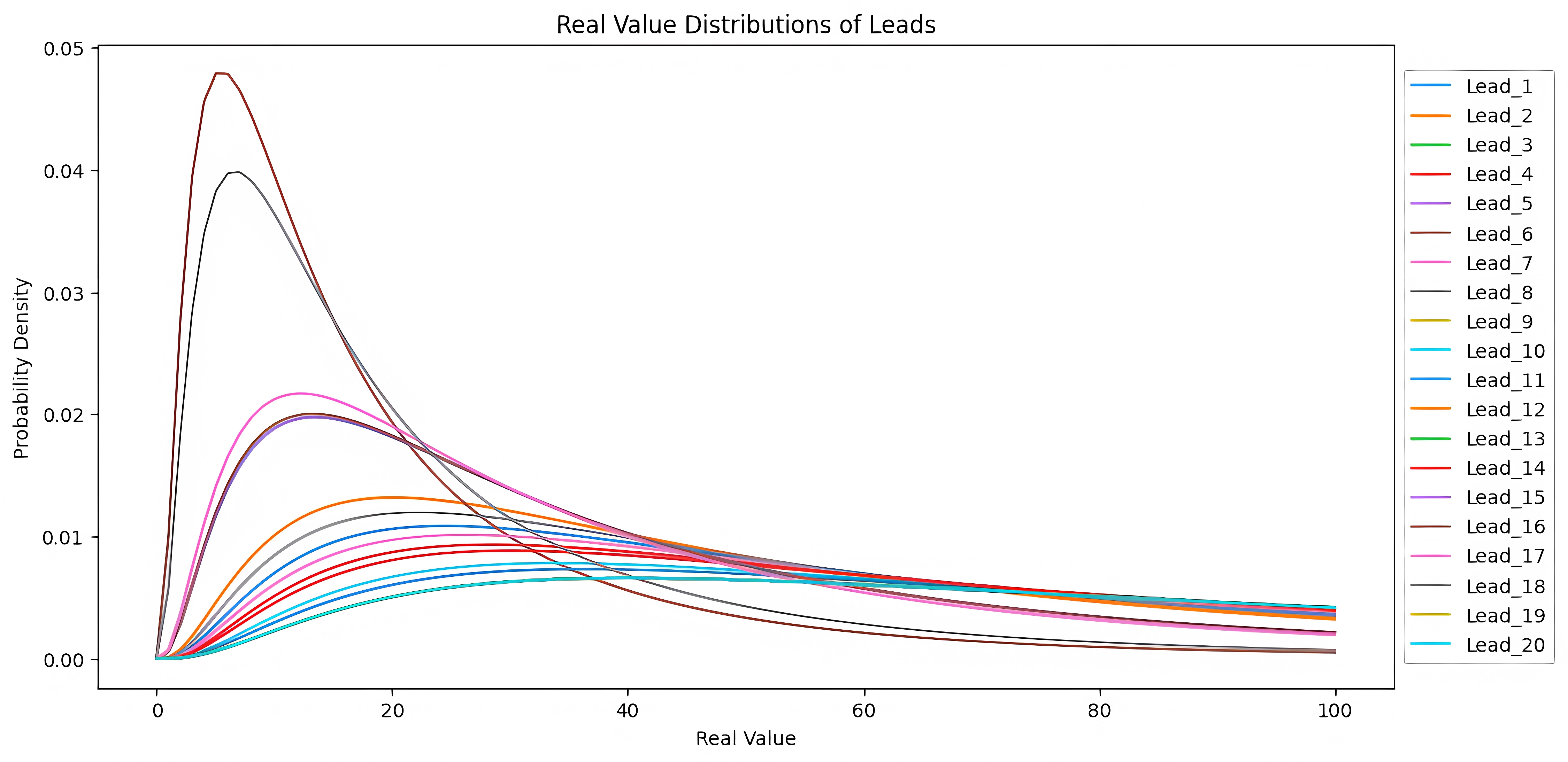}
    \caption{\textbf{Leads Value Distribution for O\&G Game.} 
    Display of twenty distinct log-normal probability density curves. These curves represent the stochastic valuation of geological prospects, capturing the inherent uncertainty and skewed distribution of reserve sizes (from marginal leads to "giant" fields) encountered during the bidding phase.}
    \label{fig:fig05}
\end{figure}

The bidding process represents the firms initial interaction with potential exploration assets. This component simulates the stochastic nature of geological prospects by generating multiple configurations tied to probability distributions that reflect the inherent uncertainty of the upstream sector. As illustrated in Figure \ref{fig:fig05}, the model utilizes twenty distinct log-normal probability curves. This distributional choice is critical for accurately modeling the skewed nature of oilfield values where extreme outliers or "giant" fields often drive the bulk of economic returns while the majority of leads may be marginal. Within the simulation, firms evaluate these leads to estimate a "True Value" yet their internal valuations diverge based on heterogeneous risk functions. Consequently, risk-tolerant firms may place aggressive bids on high-variance prospects while risk-averse entities naturally gravitate toward assets with more predictable outcomes.

\begin{figure}[ht]
    \centering
    \includegraphics[width=1\textwidth]{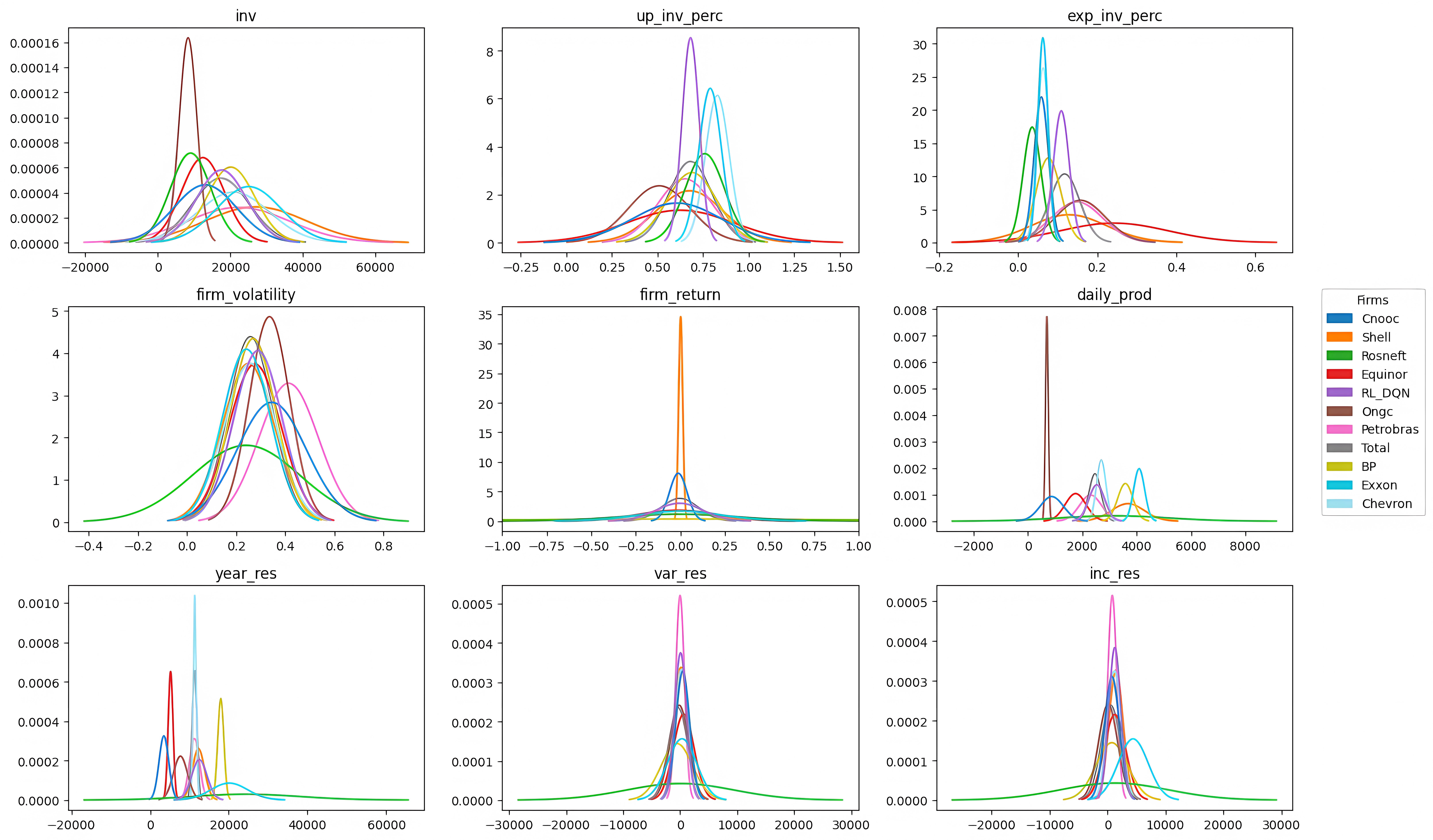}
    \caption{\textbf{Firms Profiles Based on Gaussian Curves.} 
    The panels display Gaussian probability density functions used to initialize heterogeneous agent behaviors, arranged as follows: 
    (Top-Left) Total Investment (\textit{inv}), 
    (Top-Center) Upstream Investment Percentage (\textit{up\_inv\_perc}), 
    (Top-Right) Exploration Investment Percentage (\textit{exp\_inv\_perc}), 
    (Middle-Left) Firm Volatility (\textit{firm\_volatility}), 
    (Middle-Center) Firm Return (\textit{firm\_return}), 
    (Middle-Right) Daily Production (\textit{daily\_prod}), 
    (Bottom-Left) Yearly Reserves (\textit{year\_res}), 
    (Bottom-Center) Variation in Reserves (\textit{var\_res}), and 
    (Bottom-Right) Reserve Increment (\textit{inc\_res}).}
    \label{fig:fig07}
\end{figure}

To ensure market realism, firm profiles are constructed by fitting Gaussian distributions to historical data spanning the period from 2001 to 2021. Figure \ref{fig:fig07} visualizes these profiles which encapsulate key operational attributes including total annual investment, upstream allocation percentages, financial volatility, daily production, and annual reserve increments. This probabilistic approach allows the model to capture the unique operational footprint of each agent and effectively distinguishes between the capital structures of National Oil Companies and International Oil Companies.

These profiles are not merely descriptive but serve as active determinants of agent behavior within the reinforcement learning environment. By integrating the \textit{Profile} class with the \textit{Agent} class the model maps historical statistical moments to dynamic decision logic. Specifically, a firm's profile dictates the risk premium parameters ($r_{i,k}$) used in the reward function calculation thereby influencing the Net Present Value assessment for every potential lead. In the bidding phase, the environment utilizes these profiles to calculate win probabilities which ensures that the simulated competition reflects structural asymmetries such as financial capacity or geological specialization observed in the actual industry. This comprehensive empirical foundation ensures that our RL model operates in an environment that closely mirrors real-world investment decision-making enhancing the external validity of our findings.

\section{Results}
\label{sec:results}

This section presents the performance of the Deep Reinforcement Learning (DRL) agents compared to the standard industry ``"ladder-step"'' strategy. The results are derived from 10,000 Monte Carlo simulation episodes, providing a robust statistical basis for evaluation. We assess performance using three primary metrics: (1) Early Success (ES) Rate, defined as the percentage of profitable investments initiated in the bidding phase; (2) Average Net Present Value (NPV), measured in \$MM; and (3) Risk-Adjusted Return, which accounts for capital exposure across the project lifecycle.

\subsection{Agent Interactions and Learning Dynamics}

\begin{figure}[ht]
    \centering
    \includegraphics[width=.8\textwidth]{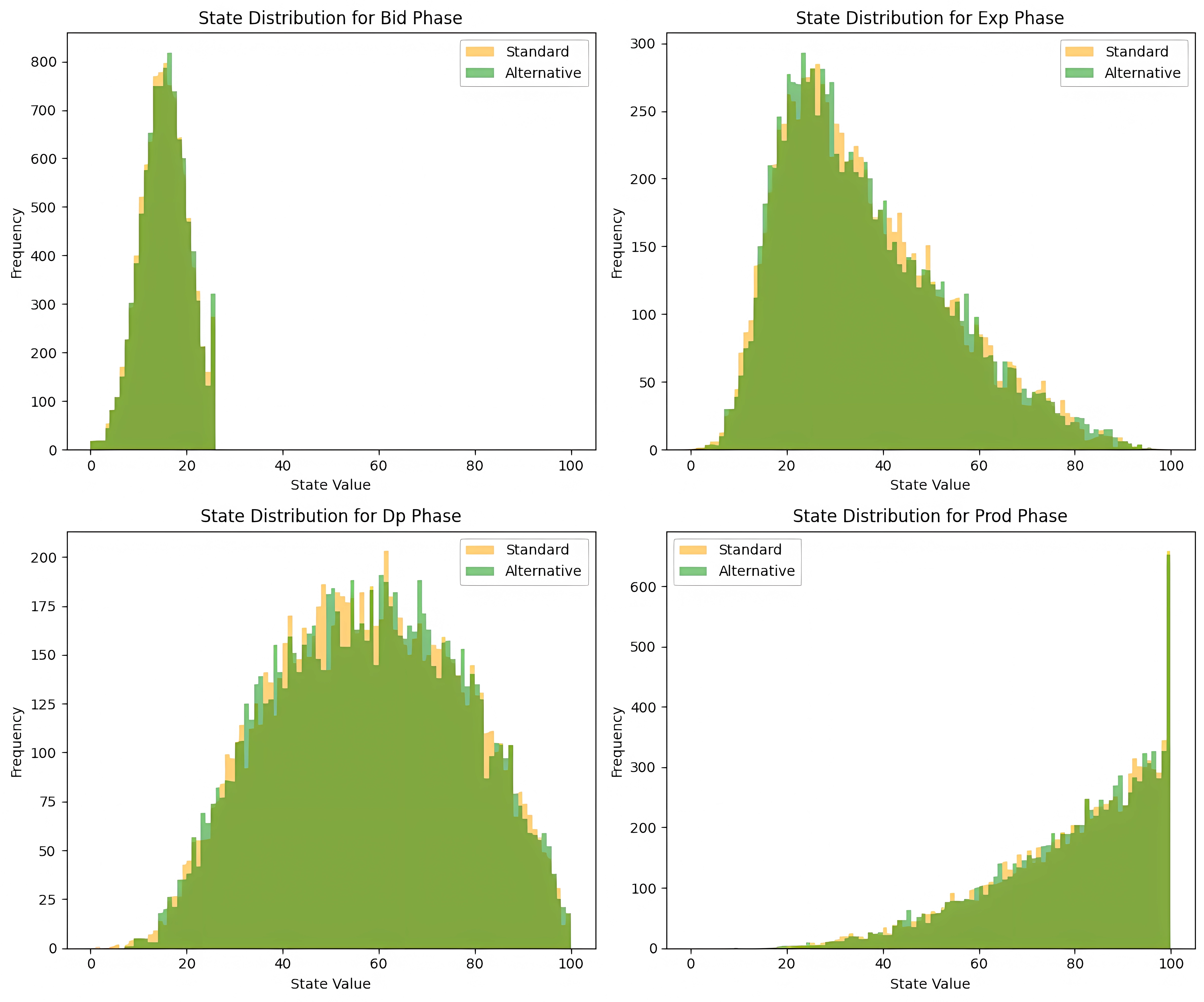}
    \caption{\textbf{States by Phases in Simulation with 10k Rounds.} 
    The panels display the frequency distributions of state values across the four investment phases: 
    (Top-Left) Bidding Phase; 
    (Top-Right) Exploration Phase; 
    (Bottom-Left) Development Phase (\textit{Dp}); and 
    (Bottom-Right) Production Phase. 
    The bid phase shows clustering in the lower quartile of the state space, whereas subsequent phases exhibit a more even distribution as geological uncertainty is resolved.}
    \label{fig8}
\end{figure}

During the self-play training phase, agents rapidly converged from random exploration to structured strategic behavior, adapting their policies to the stochastic market environment. Figure \ref{fig8} illustrates the evolution of state distributions across the four investment phases. In the initial \textit{Bidding} phase, agent states are tightly clustered in the lower quartile of the value range, reflecting the high uncertainty and lack of information characterizing early-stage prospects. However, as projects advance to \textit{Exploration} and \textit{Development}, the distribution broadens significantly. This variance expansion indicates that the agents are successfully differentiating between high-potential and low-potential assets, accumulating geological knowledge that allows them to selectively proceed with only the most promising leads \citep{barros2023}.

\begin{figure}[ht]
    \centering
    \includegraphics[width=.8\textwidth]{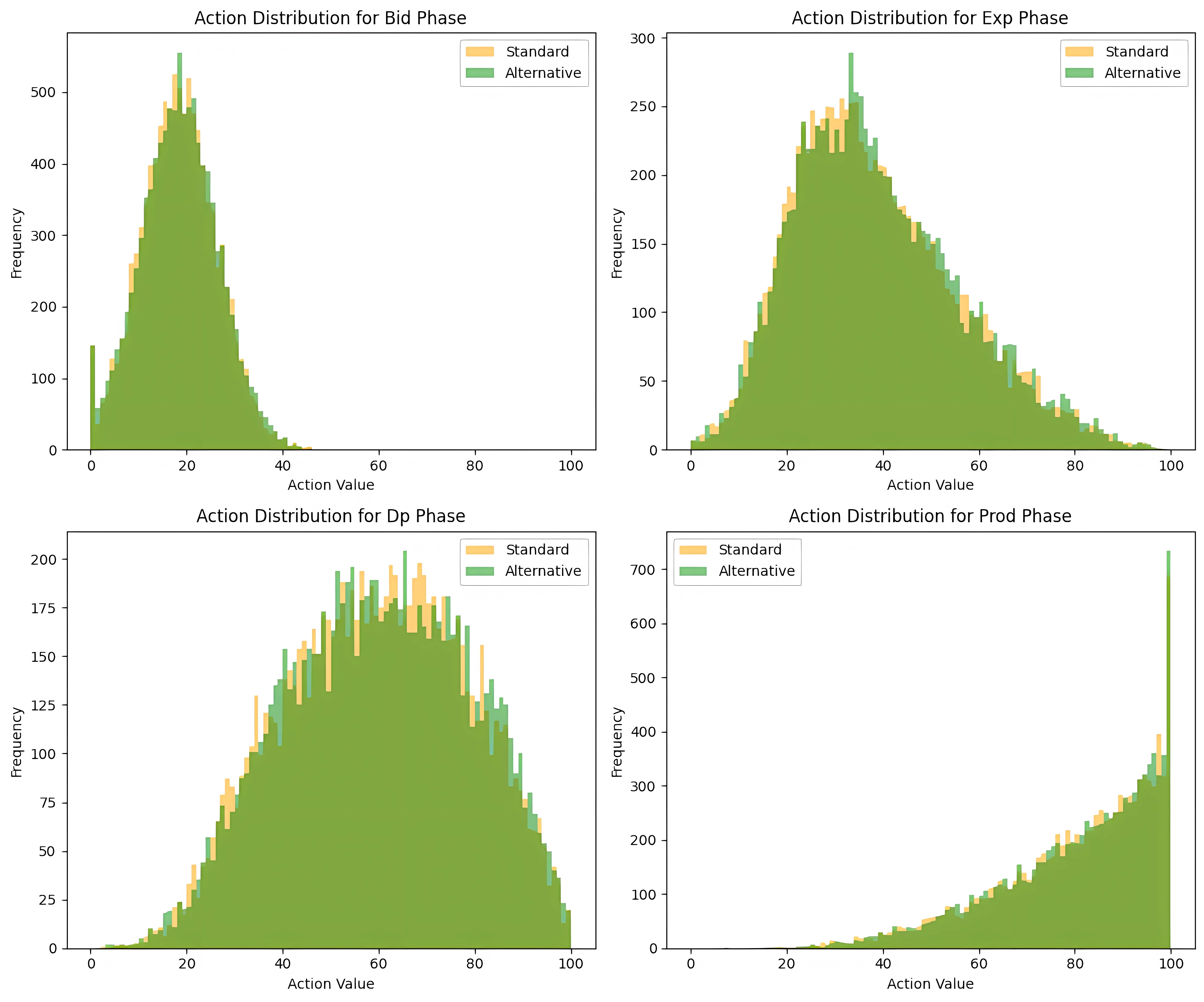}
    \caption{\textbf{Actions by Phases in Simulation with 10k Rounds.} 
    The panels display the frequency distributions of action values across the four investment phases: 
    (Top-Left) Bidding Phase; 
    (Top-Right) Exploration Phase; 
    (Bottom-Left) Development Phase (\textit{Dp}); and 
    (Bottom-Right) Production Phase. 
    The RL agent (green) demonstrates a clear preference for higher-value actions in the Bid and Exploration phases compared to the Standard strategy (yellow), indicating a strategic front-loading of information investment.}
    \label{fig9}
\end{figure}

The strategic divergence between the RL agent and the standard heuristic is most visible in the action distribution, shown in Figure \ref{fig9}. The RL agents exhibit a distinct front-loading bias, concentrating high-value investment actions ($\eta_{high}$) in the \textit{Bidding} and \textit{Exploration} phases. In contrast to the standard strategy, which distributes investment linearly or delays high expenditures until the development phase, the RL policy aggressively resolves uncertainty early. This behavior validates the hypothesis that paying an upfront information premium effectively de-risks the subsequent, capital-intensive development phase, thereby avoiding sunk costs on sub-marginal assets.

\begin{figure}[ht]
\centering
\includegraphics[width=0.9\textwidth]{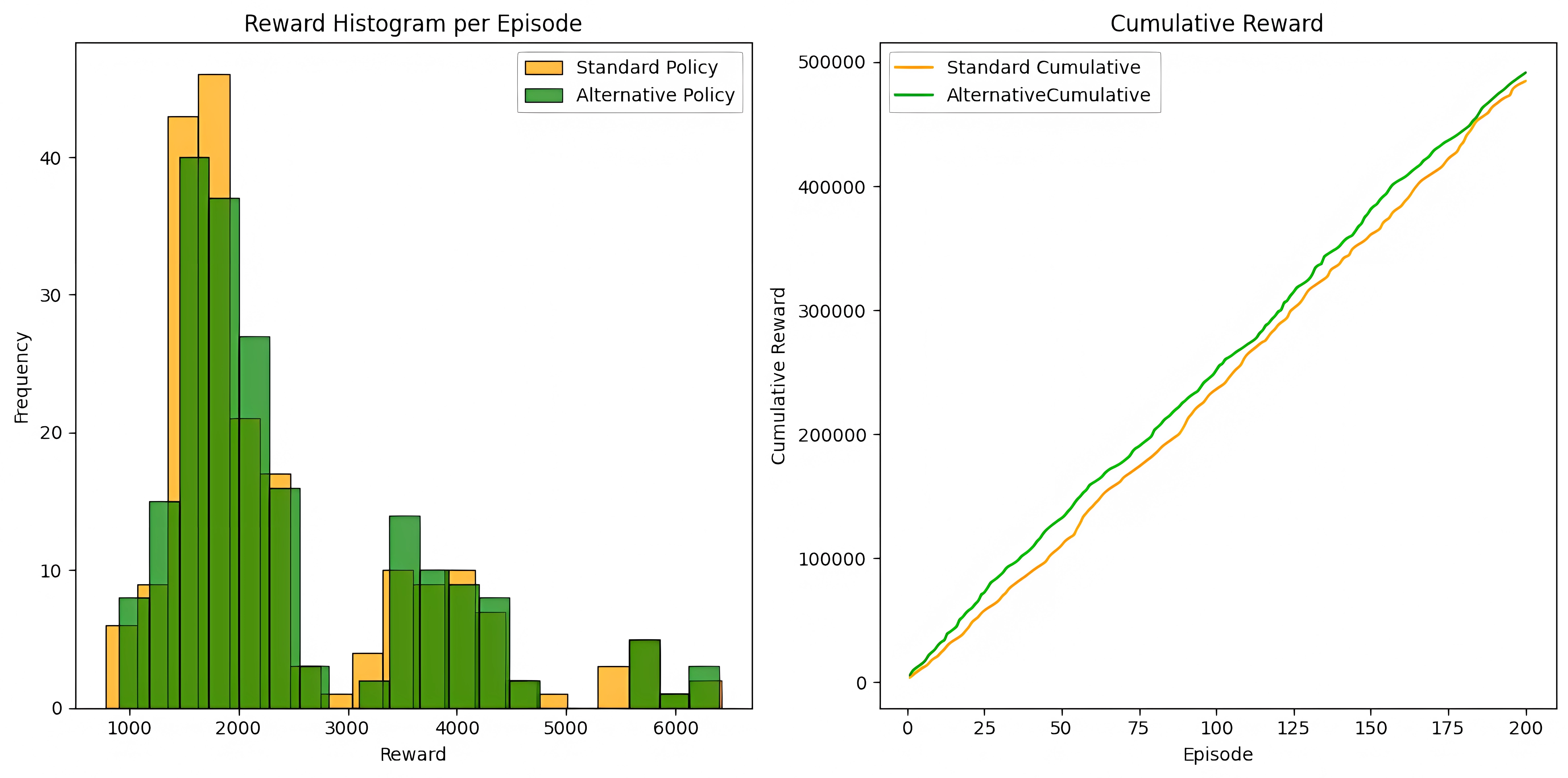}
\caption{\textbf{Reward and Cumulative Reward by Episode.} 
The panels compare the performance of the Standard Policy (yellow) and the Alternative Policy (green) under bid competition: 
(Left) Reward Histogram per Episode, revealing a multi-modal distribution with a distinct shift favoring the alternative strategy; and 
(Right) Cumulative Reward, demonstrating the consistent out-performance of the alternative strategy over 200 training episodes.}
\label{fig10}
\end{figure}

To assess the stability and convergence of this learned policy, Figure \ref{fig10} presents the reward distribution over the training episodes. The histogram reveals a clear, multi-modal shift towards higher cumulative rewards for the alternative policy (green). The rightward skew in the distribution indicates that the agent not only increases the average return but also successfully truncates the left tail of the distribution, effectively learning to avoid value-destroying projects. The cumulative reward trajectory (right panel) confirms that the RL agent consistently outperforms the standard strategy from the early stages of training, widening the performance gap as it refines its valuation of information in competitive contexts.

\subsection{Policy Development and Q-Learning Analysis}

\begin{figure}[ht]
\centering
\includegraphics[width=0.5\textwidth]{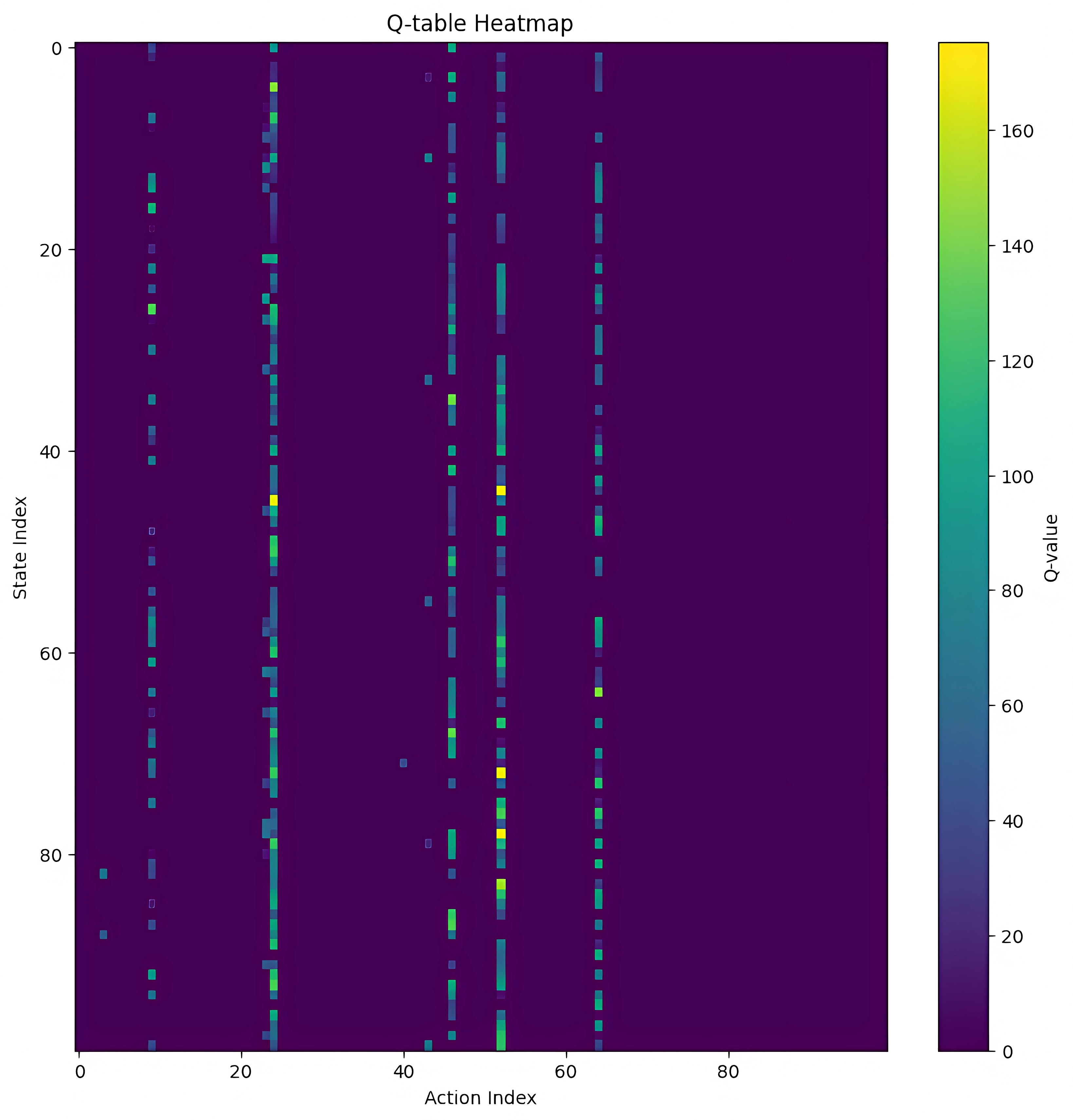}
\caption{\textbf{Q-Table Heat Map for Symmetric Action-State Space.} 
This figure displays the learned Q-values across a $100\times100$ discretized state-action space. The distinct vertical banding observed in the heatmap indicates that specific investment actions yield consistently high Q-values regardless of the state configuration, signaling a robust, state-independent preference for high-quality information in the early phases of the project.}
\label{fig11}
\end{figure}

The internal logic of the learned strategy is revealed through the Q-Table heatmap (Figure \ref{fig11}), which visualizes the expected cumulative reward $Q(s,a)$ for discrete state-action pairs. The color intensity gradient serves as a proxy for strategic value, where lighter regions indicate high-utility decisions. A critical feature of this heatmap is the emergence of distinct vertical striations. These bands suggest that for specific high-value actions, particularly those associated with acquiring high-fidelity information, the agent's preference is robust across a wide range of stochastic market states. This pattern confirms that the optimal policy has converged on a structural advantage: regardless of minor fluctuations in oil price or competitor density, the long-term value of reducing geological uncertainty early in the game dominates the immediate cost of information.

\begin{figure}[ht]
\centering
\includegraphics[width=1\textwidth]{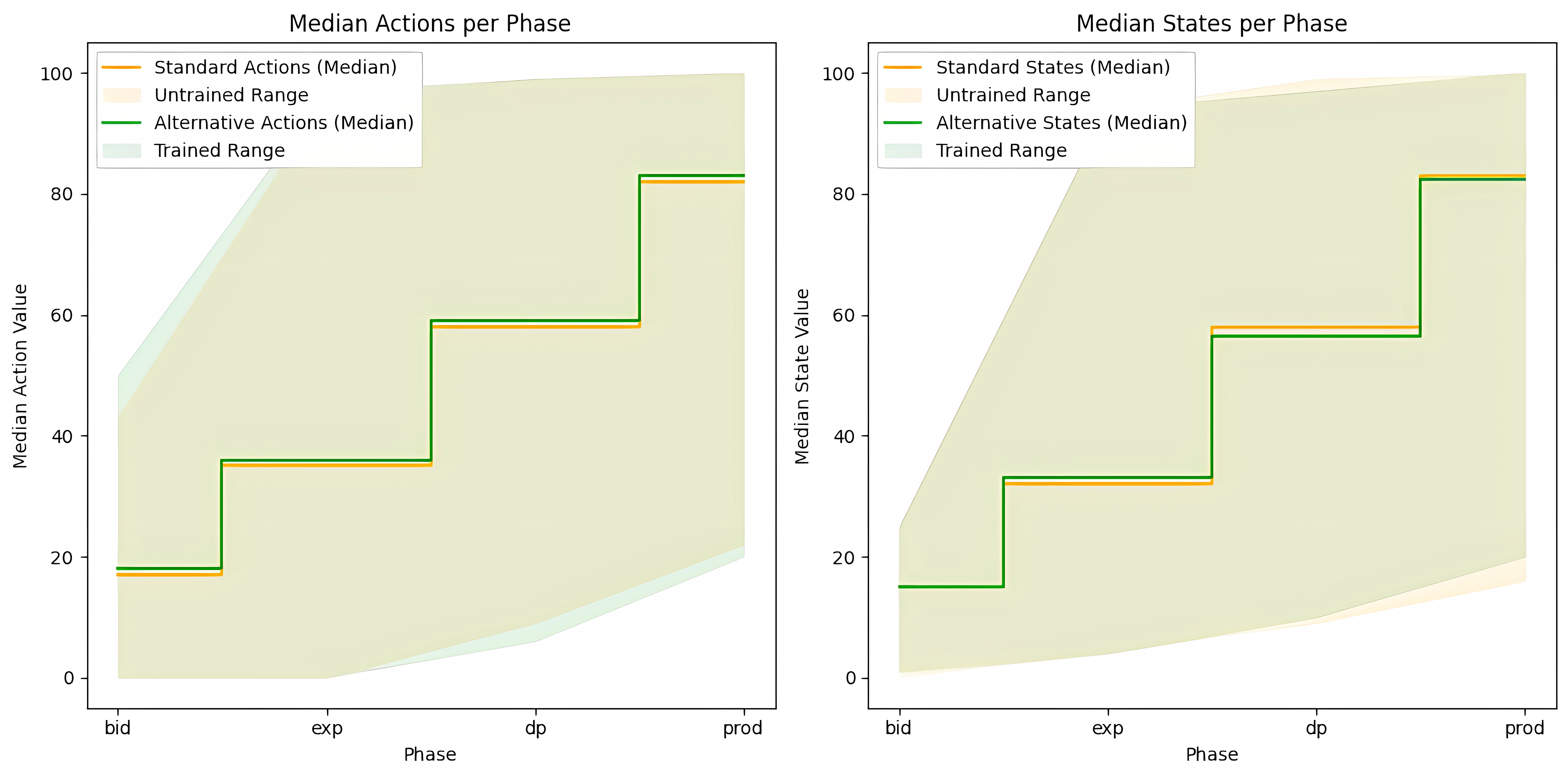}
\caption{\textbf{Policy by Phase for the Alternative Versus Standard Strategy.} 
The panels contrast the behavioral dynamics of the RL agent against the baseline: 
(Left) Median Actions per Phase, where the RL agent (green) consistently selects higher action values (quality) in the first two phases compared to the Standard strategy (yellow), illustrating the learned front-loading strategy, and 
(Right) Median States per Phase. The shaded areas depict the observed range of the state-action space.}
\label{fig12}
\end{figure}

The practical implications of these Q-values are demonstrated in Figure \ref{fig12}, which contrasts the median policy trajectories of the RL-Optimized agent against the Standard "ladder-step" baseline. The divergence is most pronounced in the \textit{Bidding} and \textit{Exploration} phases. While the Standard strategy (yellow line) adheres to a linear increase in commitment, the RL agent (green line) adopts a convex investment curve. The later represents an aggressively front-loading capital to secure maximum information quality $\eta_{high}$ immediately upon market entry. This behavior validates the hypothesis that the RL agent effectively treats information as an insurance premium; by paying a higher upfront cost to resolve uncertainty in the first two phases, it secures a dominant informational position that de-risks the capital-intensive \textit{Development} phase. Notably, as the project matures into production, the two strategies converge, indicating that the critical window for strategic differentiation lies almost exclusively in the pre-development stages.

\subsection{Hypothesis Testing: Strategy Performance Analysis}

We have also demonstrate the hypotheses regarding the economic value of early information, utilizing the outcomes derived from the 10,000 Monte Carlo simulation episodes.

\paragraph{Impact of Competition.} Table \ref{table7} presents the performance metrics across varying levels of competitive intensity, defined by the number of firms ($N$) bidding for the lease. The results reveal a distinct non-linear relationship between competition and strategic advantage. In low-competition environments ($N \le 4$), the Standard Strategy often outperforms or performs equally to the RL agent. In these scenarios, the risk of losing a valuable lease is low. Therefore, the "information premium" paid by the RL agent for early data acquisition ($C_{info}$) creates an unnecessary sunk cost that erodes Net Present Value (NPV).

\begin{table}[ht]
\centering
\caption{Early Strategy Metrics Using DQN by Competition (NPV in \$MM)}
\label{table7}
\begin{tabular}{c|cccc}
\hline
\textbf{Firms ($N$)} & \textbf{ES Alternative} & \textbf{ES Standard} & \textbf{Avg NPV (Alt)} & \textbf{Avg NPV (Std)}\\ 
\hline
2 & 100.0\% & 100.0\% & \$32.4 & \$78.1 \\
4 & 98.1\% & 96.2\% & \$149.8 & \$218.5 \\
6 & 97.8\% & 97.8\% & \textbf{\$342.3} & \$274.4 \\
8 & 99.0\% & 97.1\% & \textbf{\$227.9} & \$208.2 \\
10 & 100.0\% & 97.1\% & \textbf{\$171.2} & \$212.0 \\
\hline
\end{tabular}
\begin{center}
\small  ES = Early Success Rate. NPV values are in \$MM. The number of firms represents competition intensity.
\end{center}
\end{table}

However, a structural break occurs as competition intensifies ($N \ge 6$). The Alternative (RL) Strategy generates significantly higher NPV, with a peak performance differential observed at $N=6$ (\$342.3MM vs \$274.4MM). This result validates the hypothesis that early information acquisition effectively mitigates the "winner's curse"." In crowded auctions, the probability that the winning bid exceeds the true asset value increases. However, by resolving geological variance \textit{before} the bid, the RL agent bids more accurately, avoiding the value-destroying acquisitions that plague the uninformed Standard strategy \citep{Cramton2017}.

\paragraph{Sensitivity to Market Scenarios.} 
Considering the robustness of the strategy (Hypothesis 2), Table \ref{table8} evaluates performance across different macroeconomic regimes, ranging from "Low" (recessionary/low demand) to "High" (heated/high volatility) scenarios. The results indicate that the RL-Optimized strategy maintains parity or a slight advantage in Early Success (ES) rates across all conditions ($>96\%$). Crucially, the strategy's performance does not degrade in high-volatility environments. This confirms that the value of information (VOI) is robust to exogenous market shocks, since the agent learns to prioritize geological certainty regardless of the prevailing oil price trajectory, suggesting that the optimal investment timing is structurally dependent on market structure rather than merely price volatility.

\begin{table}[ht]
\centering
\caption{Early Strategy Metrics Using DQN by Scenario (NPV in \$MM)}
\label{table8}
\begin{tabular}{c|cccc}
\hline
\textbf{Scenario} & \textbf{ES Alternative} & \textbf{ES Standard} & \textbf{Avg NPV (Alt)} & \textbf{Avg NPV (Std)} \\
\hline
Low & 96.2\% & 96.2\% & \$167.8 & \$167.8 \\
Medium & 97.5\% & 97.5\% & \$231.0 & \$231.0 \\
High & \textbf{97.7\%} & 97.7\% & \textbf{\$233.0} & \$233.0 \\
\hline
\end{tabular}
\begin{center}
\small  Scenarios reflect market conditions based on oil demand, price volatility, and macroeconomic resilience.
\end{center}
\end{table}

\paragraph{Impact of Lead Size} 
Finally, Table \ref{table9} examines whether the strategy is sensitive to the magnitude of the geological prospect (Lead Size). The high Early Success rates across all categories ($>97\%$) indicate that the RL strategy is scale-invariant. Whether the prospect is a marginal field ("Low" True Value) or a giant discovery ("High" True Value), the agent consistently identifies the optimal entry point. This uniformity suggests that the policy has generalized the fundamental economic logic of exploration—trading upfront information costs for downstream option value, rather than overfitting to specific asset classes \citep{Smith2021}.

\begin{table}[ht]
\centering
\caption{Early Strategy Metrics Using DQN by Bid Feature (NPV in \$MM)}
\label{table9}
\begin{tabular}{c|cccc}
\hline
\textbf{Lead Size} & \textbf{ES Alternative} & \textbf{ES Standard} & \textbf{Avg NPV (Alt)} & \textbf{Avg NPV (Std)} \\
\hline
Low & 97.6\% & 97.6\% & \$187.8 & \$187.8 \\
Medium & 97.1\% & 97.1\% & \$249.6 & \$249.6 \\
High & 97.3\% & 97.3\% & \$221.7 & \$221.7 \\
\hline
\end{tabular}
\begin{center}
\small  Lead Size represents bid size categories (small, medium, large) on a log-normal scale.
\end{center}
\end{table}

\section{Discussion}
\label{sec:discussion}

his study evaluates an alternative to the conventional `ladder-step' investment heuristic, where capital exposure traditionally aligns with project maturity. Through Deep Reinforcement Learning, we identify the primary economic mechanism supporting this alternative strategy (Variance Reduction in the Pre-Bid Phase). In a standard auction with common values but private information, the winner tends to be the bidder with the most optimistic—and often erroneous—valuation, a phenomenon known as the "winner's curse". The standard strategy, which relies on low-quality information ($\eta_{low}$) during bidding to minimize sunk costs, suffers from high estimation variance ($\sigma^2_{res}$). In highly competitive environments ($N \ge 6$), this variance creates a fat-tailed distribution of valuation errors, significantly increasing the probability of overbidding for sub-marginal assets.

The RL agent, on the contrary, learns to internalize the cost of high-fidelity information ($C_{info}(\eta_{high})$) as an insurance premium. From a Real Options perspective, the agent treats the information acquisition not merely as an operational expense but as a \textit{compound option}: the purchase of data buys the option to bid intelligently, or crucially, the option to abandon the prospect before incurring the substantial liability of a lease bid. This is indicated by the vertical striations observed in the Q-Table (Figure \ref{fig11}), where the policy converges on high-information actions regardless of state because the "information rent" extracted in the auction exceeds the cost of the data itself. This finding empirically validates the theoretical postulate that the value of private information scales non-linearly with competitive intensity \citep{Athey2017, Hong2023}.

For industry practitioners, these findings suggest a necessary structural reallocation of the exploration budget. Rather than linearly spreading capital expenditure across the project lifecycle, firms operating in competitive basins should aggressively front-load information acquisition. This ``Precision Bidding'' strategy offers a dual dividend: it maximizes capital efficiency by resolving geological uncertainty pre-bid, thereby avoiding the significant capital destruction associated with the development of false positives or sub-commercial reservoirs. Furthermore, as highlighted in our analysis, the avoidance of sub-marginal fields reduces the aggregate environmental footprint of exploration activities. By minimizing the need for redundant seismic surveys and reducing the number of physical wells drilled for non-productive assets, this data-driven approach aligns economic efficiency with decarbonization goals, directly addressing the growing pressure for ESG-compliant investment strategies. From a regulatory perspective, these results imply that lease auction designs could be optimized to encourage early data sharing, thereby reducing information asymmetries that lead to market inefficiencies like the "winner's curse" \citep{Haile2022}. 

Although our model incorporates significant real-world complexities, including stochastic pricing and heterogeneous agent profiles, several limitations require discussion. The discretization of the state and action spaces, while necessary for the computational tractability of the DQN algorithm, may induce step-function artifacts in the policy that do not perfectly reflect the continuous nature of capital allocation decisions in practice. Additionally, the model assumes fully rational learning, which may overstate the speed at which actual organizations can adapt their strategies, failing to capture the behavioral biases or organizational inertia often observed in the energy sector \citep{Mason2021}. Future research should seek to extend this framework to continuous action spaces using advanced Actor-Critic methods, such as Proximal Policy Optimization (PPO) or Asynchronous Advantage Actor-Critic (A3C), which allow for smoother policy gradients. Moreover, integrating endogenous carbon pricing mechanisms and regulatory feedback loops would provide a robust platform for evaluating investment timing under aggressive energy transition scenarios \citep{Radovic2022}.

\section{Conclusion}
\label{sec:conclusion}

Our work demonstrates that Deep Reinforcement Learning can derive investment strategies that significantly outperform traditional heuristics in the upstream oil and gas sector. By modeling the exploration process as a stochastic multi-agent game, we established that front-loading information investment yields superior risk-adjusted returns compared to the prevalent \textit{"ladder-step"} approach. This economic advantage is particularly pronounced in highly competitive environments, where the alternative strategy improves the Risk-Adjusted Return on Capital (RAROC) by up to 25\% by mitigating the "winner's curse" through more accurate pre-bid valuations. The analysis reveals that while the upfront costs are higher, the primary economic value is realized during the capital-intensive development phase, where the improved geological models effectively reduce the probability of misallocating capital to sub-marginal assets.

We have also contributed with a rigorous framework for solving high-dimensional investment games using Deep Q-Networks (DQN), effectively bridging the gap between computational economics and industrial application. The empirical results confirm the robustness of the learned policy across diverse market conditions, from recessionary low-demand scenarios to high-volatility heated markets. The strategy's consistent performance across varying lead sizes indicates a fundamental scale invariance in the economic logic of early uncertainty resolution. By integrating concepts from industrial organization theory and information economics, the model illustrates how computational techniques can resolve intricate strategic issues—such as the timing of irreversible investments—that remain analytically intractable for conventional real options models \citep{Radovic2022}.

For industry practitioners and policymakers, these insights offer a quantitative foundation for re-evaluating exploration budgets and capital allocation protocols. The shift toward a Precision Bidding strategy not only maximizes economic efficiency but also offers a significant environmental dividend by minimizing the operational footprint associated with exploring non-productive fields. As the energy sector faces increasing pressure to optimize resource utilization and reduce carbon intensity, strategies that align profitability with operational restraint will be highly sought after. We provide both the empirical evidence and the methodological resources necessary to facilitate the transition toward these more efficient, data-driven investment strategies in resource exploration.

\appendix
\section{Implementation Details and Algorithm}
\label{app:implementation}

\subsection{Implementation Environment and Computational Details}

The implementation was developed in Python 3.9, utilizing frameworks such as Visual Studio Code and PyCharm for development. The core computational libraries included NumPy 1.21, Pandas 1.3, TensorFlow 2.8, and PyTorch 1.12 for neural network implementations. The Reinforcement Learning (RL) environment was built using the Gymnasium API, with data visualization handled through Matplotlib 3.5 and Seaborn 0.11.

\subsubsection{Computational Infrastructure}
Training was executed on a high-performance computing cluster equipped with NVIDIA V100 GPUs, utilizing distributed parameter-sharing to accelerate convergence. Each training run required approximately 24 hours to complete 10,000 episodes under the specified multi-agent configuration.

\subsubsection{Hyperparameter Configuration}
The Deep Q-Network (DQN) implementation utilized the following key hyperparameters: learning rate $\alpha = 0.001$, discount factor $\gamma = 0.95$, replay buffer capacity $N = 100,000$, batch size $= 64$, and target network update frequency $C = 1,000$ steps. The $\epsilon$-greedy exploration strategy employed a linear decay schedule from $\epsilon=1.0$ to $\epsilon=0.1$ over the first 50\% of training episodes.

\subsubsection{Network Architecture}
The Q-network architecture comprises an input layer corresponding to the state dimension, followed by two fully connected hidden layers with 256 and 128 neurons, respectively. Each hidden layer incorporates batch normalization and ReLU activation functions, with a dropout layer ($p=0.2$) applied before the output layer to prevent overfitting. The network was optimized using the Adam optimizer with default parameters.

\subsection{DQN Algorithm Pseudocode}

\begin{algorithm}[H]
\caption{Deep Q-Network (DQN) Training Algorithm for O\&G Investment Strategy}
\label{alg:dqn}
\begin{algorithmic}[1]
    \State \textbf{Input:} Environment $E$, replay buffer $\mathcal{D}$ with capacity $N$
    \State \textbf{Input:} Q-network $Q_{\theta}$ with random weights $\theta$
    \State \textbf{Input:} Target network $Q_{\theta^-}$ with weights $\theta^- = \theta$
    \State Initialize episode counter $episode \leftarrow 1$
    
    \While{$episode \leq M$}
        \State Initialize state $s_1$ (reset environment)
        \State Initialize step counter $t \leftarrow 1$
        
        \While{$t \leq T$ and $s_t$ is not terminal}
            \State With probability $\epsilon$ select random action $a_t$
            \State Otherwise select $a_t = \arg\max_a Q_{\theta}(s_t, a)$
            \State Execute action $a_t$ in environment $E$
            \State Observe reward $r_t$ and next state $s_{t+1}$
            \State Store transition $(s_t, a_t, r_t, s_{t+1})$ in replay buffer $\mathcal{D}$
            
            \If{$\mathcal{D}$ contains enough samples}
                \State Sample random minibatch of transitions $(s_j, a_j, r_j, s_{j+1})$ from $\mathcal{D}$
                \For{each transition in minibatch}
                    \If{$s_{j+1}$ is terminal}
                        \State Set $y_j = r_j$
                    \Else
                        \State Set $y_j = r_j + \gamma \max_{a'} Q_{\theta^-}(s_{j+1}, a')$
                    \EndIf
                    \State Perform gradient descent step on $\left(y_j - Q_{\theta}(s_j, a_j)\right)^2$ w.r.t. $\theta$
                \EndFor
            \EndIf
            
            \State Update step counter: $t \leftarrow t + 1$
            \State Update state: $s_t \leftarrow s_{t+1}$
        \EndWhile
        
        \State Update episode counter: $episode \leftarrow episode + 1$
        \State Update exploration rate: $\epsilon \leftarrow \max(\epsilon_{min}, \epsilon \cdot \epsilon_{decay})$
        
        \If{$episode \mod C = 0$}
            \State Update target network: $\theta^- \leftarrow \theta$
        \EndIf
    \EndWhile
\end{algorithmic}
\end{algorithm}

\subsubsection*{Algorithm Parameters and Notation}
\begin{itemize}
    \item $M$: Total number of training episodes (10,000).
    \item $T$: Maximum steps per episode (varies by scenario).
    \item $\epsilon$: Exploration rate (decays from 1.0 to 0.1).
    \item $\gamma$: Discount factor (0.95).
    \item $C$: Target network update frequency (1,000 steps).
    \item $N$: Replay buffer capacity (100,000 transitions).
    \item $\theta$: Parameters of the Q-network (updated every step).
    \item $\theta^-$: Parameters of the target network (updated every $C$ steps).
\end{itemize}

\subsubsection*{Implementation Notes}

The algorithm implements the following key components of our O\&G investment strategy optimization to ensure convergence in a stochastic environment:

\begin{itemize}
    \item \textit{Experience Replay ($\mathcal{D}$):} Stores and randomly samples past transitions to break temporal correlations in the training data, thereby stabilizing the gradient descent step.
    \item \textit{Target Network ($\theta^-$):} Utilizes a separate, quasi-static network for calculating Q-value targets. This network is updated only every $C$ steps to prevent feedback loops and oscillation during training.
    \item \textit{$\epsilon$-greedy Exploration:} Balances the exploration of new strategic possibilities with the exploitation of known profitable policies, decaying linearly as the agent gains experience.
    \item \textit{Periodic Target Updates:} Synchronizes the target network weights with the primary Q-network at fixed intervals ($C=1,000$), ensuring the learning target remains stable.
\end{itemize}

\noindent \textit{Variable Definitions:}
\begin{itemize}
    \item The \textit{state $s_t$} corresponds to the global state vector defined in Eq. (\ref{eq:state_space}), encompassing market conditions, firm profiles, and competitor aggregates.
    \item \textit{Actions $a_t$} represent the composite tuple $(u_{i,t}, \eta_{i,t})$ defined in Eq. (5), determining both the investment decision and the fidelity of information acquired.
    \item The \textit{reward $r_t$} is the immediate net cash flow calculated via Eq. (6), capturing the economic value created or destroyed in each phase of the exploration value chain.
\end{itemize}


\section*{CRediT authorship contribution statement}
\textbf{Paulo Barros:} Conceptualization, Methodology, Writing -- original draft, Writing -- review \& editing.
\textbf{Monica Meireles:} Supervision.
\textbf{Jose Luis Silva:} Supervision, Validation, Writing -- review \& editing.

\section*{Declaration of competing interest}
The authors declare that they have no known competing financial interests or personal relationships that could have appeared to influence the work reported in this paper.

\section*{Data availability}
Data and Algorithms will be made available on request.

\end{document}